\documentclass[reprint,amsmath,floatfix,citeautoscript]{revtex4-1}
\usepackage{dsfont}
\usepackage{tabularx}
\usepackage{booktabs}
\AtBeginDocument{
\heavyrulewidth=.08em
\lightrulewidth=.05em
\cmidrulewidth=.03em
\belowrulesep=.65ex
\belowbottomsep=0pt
\aboverulesep=.4ex
\abovetopsep=0pt
\cmidrulesep=\doublerulesep
\cmidrulekern=.5em
\defaultaddspace=.5em
}
\newcolumntype{C}[1]{>{\centering\arraybackslash}p{#1}}
\usepackage{siunitx}
\usepackage{graphicx}
\usepackage{xcolor}
\usepackage[colorlinks,urlcolor=blue]{hyperref}
\setcitestyle{super}

\begin{document}

\author{Martin St\"{o}hr}
\author{Leonardo Medrano Sandonas}
\author{Alexandre Tkatchenko}
\email{alexandre.tkatchenko@uni.lu}
\affiliation{Department of Physics and Materials Science, University of Luxembourg, L-1511 Luxembourg, Luxembourg.}

\title{Accurate Many-Body Repulsive Potentials for Density-Functional Tight-Binding from Deep Tensor Neural Networks}


\begin{abstract}
\noindent
We combine density-functional tight-binding (DFTB) with deep tensor neural networks (DTNN) to maximize the strengths of both approaches in predicting structural, energetic, and vibrational molecular properties. The DTNN is used to learn a non-linear model for the localized many-body interatomic repulsive energy, which so far has been treated in an atom-pairwise manner in DFTB. Substantially improving upon standard DFTB and DTNN, the resulting DFTB-NN\textsubscript{rep} model yields accurate predictions of atomization and isomerization energies, equilibrium geometries, vibrational frequencies and dihedral rotation profiles for a large variety of organic molecules compared to the hybrid DFT-PBE0 functional. Our results highlight the high potential of combining semi-empirical electronic-structure methods with physically-motivated machine learning approaches for predicting localized many-body interactions. We conclude by discussing future advancements of the DFTB-NN\textsubscript{rep} approach that could enable chemically accurate electronic-structure calculations for systems with tens of thousands of atoms.
\end{abstract}

\maketitle

Atomistic modeling has by now become an integral part of studying and understanding systems in chemistry, biology and materials science. The two workhorse methods in that regard are Density Functional Theory (DFT) and empirical molecular mechanics force fields. With the rapidly growing interest in a microscopic understanding of systems at increasingly larger length and time scales, these two approaches keep facing considerable limitations. Despite the ever-growing availability of computational resources and high performance implementations, DFT is still limited in terms of tractable system sizes due to the associated computational workload. Molecular mechanics approaches, on the other side, are often insufficient due to the lack of access to electronic properties and limited transferability. An intermediate level of theory between the two can offer a promising alternative. This is provided by semi-empirical methods, which include an explicit quantum-mechanical treatment of electrons, thus providing access to electronic properties, while requiring only a fraction of the computational cost of DFT. Semi-empirical methods typically also offer an increased accuracy and transferability when compared to classical force fields.~\cite{thiel14}

Among others,~\cite{thiel14,sankey89,lewis01,stewart07,bannwarth19} the Density-Functional Tight-Binding (DFTB) formalism~\cite{porezag95,elstner98,gaus11} is one of the most popular representatives of this class. DFTB is based on DFT using a superposition of (confined) atomic densities, $\rho_0 = \sum \rho_A$, as an approximation to the real electron density.~\cite{porezag95} This original formulation has subsequently been extended to also allow for redistribution of electrons. This is achieved by expanding the density functional around the reference density $\rho_0$ in terms of changes in the electron distribution.~\cite{elstner98,koskinen09,gaus11} Depending on the order of this expansion we speak of DFTB2 (second order) or DFTB3 (third order). The resulting energy functional can in both cases be written as a sum of an electronic energy, $E_{\rm DFTB}^{\rm (el)}$, and the so-called repulsive energy, $E_{\rm rep}$. The electronic part of the total energy is evaluated from a tight binding Hamiltonian. Using a two-center approximation, the Hamiltonian elements can be obtained from reference DFT calculations of (confined) atoms and diatomic molecules introducing only a minimal amount of parameters.~\cite{koskinen09} The repulsive energy, on the other side, is usually obtained in a much less straightforward manner. Formally, it is defined as,
\begin{equation}
  \begin{aligned}
    E_{\rm rep}\left[\rho_0\right] =\, &E_{\rm xc}\left[\rho_0\right] - \int v_{\rm xc}\left[\rho_0\right] \rho_0\,d\mathbf{r}\\
    &+ E_{\rm nuc} - \frac{1}{2}\int V_{\rm H}\left[\rho_0\right]\rho_0\,d\mathbf{r} \:\:,
  \end{aligned}
\end{equation}
with $E_{\rm xc}$ being the exchange-correlation energy, $v_{\rm xc}$ the corresponding potential, $E_{\rm nuc}$ the nuclear repulsion energy, and $V_{\rm H}$ the Hartree potential.~\cite{koskinen09,hourahine20}
As it only depends on the reference density, $\rho_0$, and atomic positions, \textit{E}\textsubscript{rep} can further be considered a more local property. In lieu of the above formal definition, \textit{E}\textsubscript{rep} is in practice represented by atom-pairwise potentials fitted to DFT reference calculations. So,
\begin{equation}
  \begin{aligned}
    E_{\rm rep} &= E_{\rm DFT} - E_{\rm DFTB}^{\rm (el)} \qquad\\
    \mbox{and}\qquad \mathbf{F}_{\rm rep} &= \mathbf{F}_{\rm DFT} - \mathbf{F}_{\rm DFTB}^{\rm (el)}\:\:.
  \end{aligned}
  \label{eq:Erep_Frep}
\end{equation}
The parameterization of repulsive potentials thus represents a complex multidimensional fitting problem of atomic pair-potentials to DFT reference results, which renders it the most intricate step in the development of DFTB parameterizations.
Various (semi-)automated approaches have been proposed to tackle this task as of today,~\cite{knaup07,gaus09,koskinen09,bodrog11,doemer13,oliveira15,chou16,krishnapriyan17} but in practice many cases still do require inevitably subjective manual adjustments.
An optimal DFTB parameterization should finally provide access to accurate electronic as well as energetic and structural properties. Optimization of the parameters governing the electronic DFTB Hamiltonian is thereby often done separately based on reproducing electronic properties as obtained in DFT or other reference methods. The final performance for energetic and structural properties is then largely governed by the repulsive energy.

Fitting repulsive potentials according to eq.~\eqref{eq:Erep_Frep} allows DFTB to often provide results at the DFT level of accuracy, but inevitably introduces empirical contributions to the repulsive potential to correct for the approximations in the remaining formalism. These contributions also introduce a beyond-pairwise character to \textit{E}\textsubscript{rep}. Consequently, the inherent limitations of the traditional pairwise formulation have proven a major pitfall for accuracy and general validity. For example, phonon band structures have been shown to be often poorly described by traditional DFTB and highly dependent on the employed repulsive potentials.~\cite{gaus13,kamencek20,niehaus19} Independent of the formalism to treat van der Waals interactions, unit cell volumes of molecular crystals can be considerably underestimated.~\cite{mortazavi18,kamencek20} Traditional repulsive potentials without empirical fixes have also failed to correctly describe the stability of Zundel ions and associated proton transfer barriers.~\cite{goyal11} As observed for the different phases of ZnO,~\cite{hellstrom13} optimal repulsive potentials can depend strongly on the chemical environment --- a final confirmation of the breakdown of the pairwise-additive approximation.
In this regard, tailoring parameterizations for a specific purpose can alleviate the shortcomings for individual properties.\cite{goyal11,hellstrom13,niehaus19} A multitude of applications requires simultaneous accuracy in a variety of properties, however.
Together with the still considerable empiricism and human effort involved in current approaches, this calls for a more advanced description of repulsive potentials beyond the pairwise picture with a straightforward access to optimization in the corresponding high-dimensional functional space.

A very successful tool for interpolating and exploring the space of high-dimensional functions are Machine Learning (ML) approaches. Since early applications of ``learning machines'', such ML-based methods have recently gained tremendous interest in the context of modeling molecular systems and materials. Given sufficient reference data, modern ML techniques aided by chemical and physical knowledge have been proven to show great success in predicting physico-chemical properties of molecules and materials.~\cite{rajan05,montavon13,Schuett2017,butler18,anatole18} In many cases the data-driven predictions reach virtually the same level of accuracy as the provided reference data. Nowadays, several methods ranging from Neural Network (NN) potentials to Kernel Ridge Regression approaches using diverse molecular descriptors have been put forward, substantially accelerating high-accuracy atomistic modeling and advancing our understanding of chemical compound space.~\cite{behler07,bartok10,hansen15,pronobis18,Schuett2019,chmiela17,faber18} Albeit aided by chemical intuition, these approaches are still mostly based on data-driven inference instead of physical laws and typically rest on an assumption of locality.
Molecular properties including interatomic interactions, however, are governed by quantum mechanics and involve a variety of non-local characteristics and phenomena. As a result, the applicability of ML models is strongly limited due to the lack of decisive effects like long-range electrostatics or van der Waals dispersion interactions. Such phenomena and effects can be readily captured by approximate quantum-mechanical approaches, like (electronic) DFTB, or augmenting models, however.~\cite{elstner98,brandenburg14,stoehr16,mortazavi18,panosetti20}

Besides the limitations due to the locality of current ML models, data-driven prediction of global properties additionally suffers from the sheer depth and complexity of the global chemical space. The number of samples needed to infer a model for global properties of a given set of molecules scales with the number of compositional, configurational, and conformational degrees of freedom. Local (chemical) environments as required for the prediction of local properties, on the other side, span a drastically reduced and most importantly bound space. For the dataset and applications considered in this work, for example, we can estimate the number of training instances for local interactions to be three orders of magnitude smaller than to cover the global chemical space (see SI, Section~4).
A combined formalism of ML potentials for (more) localized interactions, such as the repulsive energy, and approximate electronic structure methods for non-local effects, thus, represents a very promising approach to obtain an accurate and transferable methodology for studying realistic and practically-relevant systems.
\begin{figure}[bt]
 \centering
 \includegraphics[width=\linewidth]{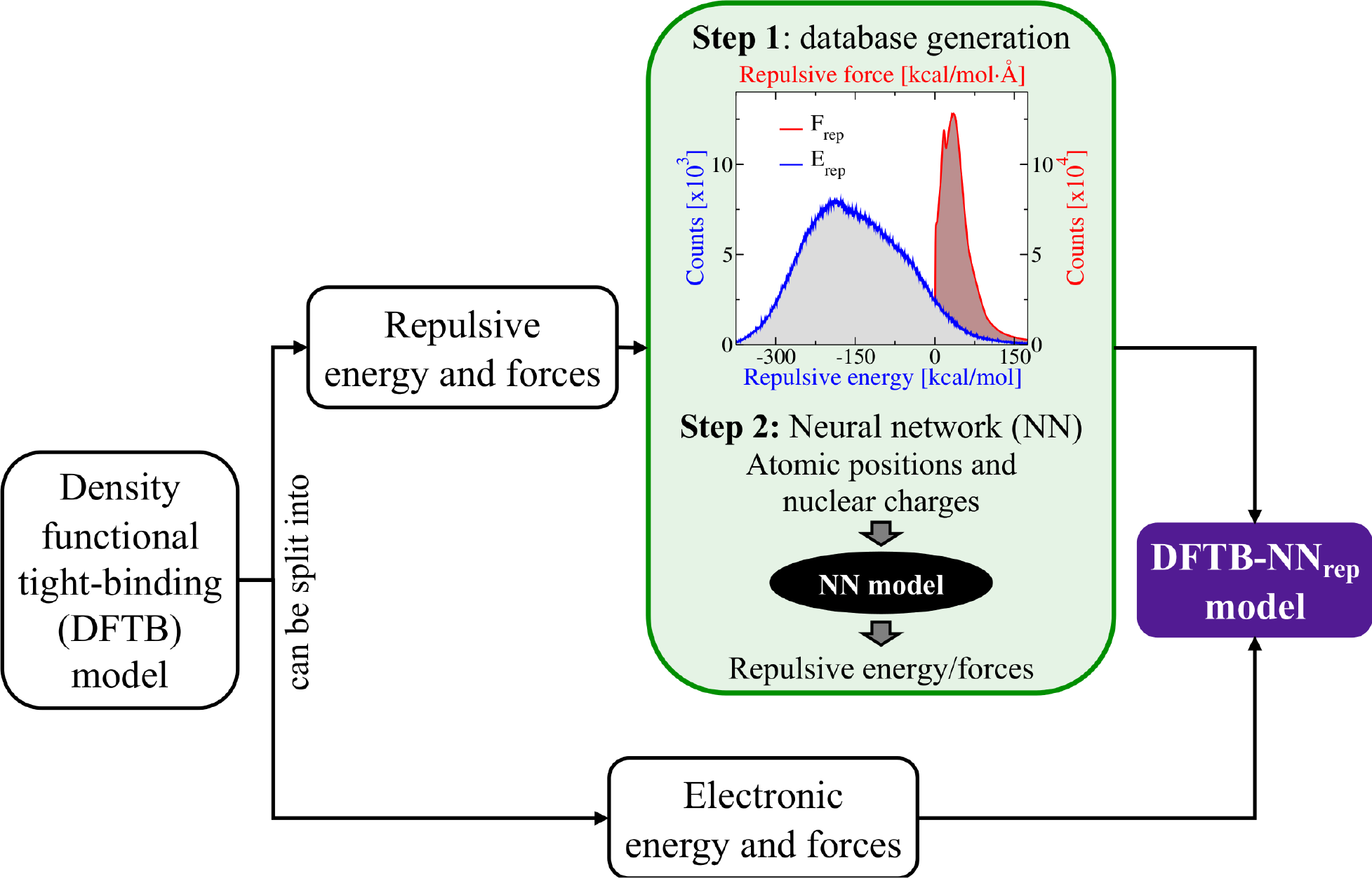}
 \caption{\textbf{Schematic representation of the DFTB-NN\textsubscript{rep} framework presented in the current work.} DFTB repulsive energies and forces are via a (deep tensor) neural network model based on PBE0 reference data, while electronic energies are properties are calculated within a Density-Functional Tight-Binding formalism.}
 \label{fig1}
\end{figure}
For the case of DFTB, such combinations have been proposed and studied recently, providing a more straightforward access to atom-pairwise repulsive potentials from Gaussian process regression~\cite{panosetti20} or introducing partial non-additivity via bond-type-dependent pair-potentials.~\cite{kranz18} In this work, we propose to further extend the latter idea to fully many-body ML potentials in order to account for beyond-pairwise repulsive contributions. This is to some extent similar in spirit to the $\Delta$-ML approach, where it is proposed to use ML models to correct the final energy of semi-empirical methods.~\cite{ramakrishnan15,zhu19} As the final DFTB energy is already fitted to reproduce DFT, such corrections can involve a rather unspecific and noisy objective quantity, however. Targeting the full repulsive potential is expected to provide a well-behaved and smooth quantity with a distinct mapping to molecular features.

To exemplify the potential and success of combining electronic structure methods with ML potentials for localized many-body interactions, we here use (electronic) DFTB in conjunction with a global Deep Tensor NN (DTNN) model. The resulting DFTB-NN\textsubscript{rep} formalism is demonstrated to provide highly accurate energetic, structural and vibrational properties for a vast range of organic molecules across their respective conformational space. All DFTB calculations have been carried out on the DFTB3~\cite{gaus11} level employing recent 3ob parameters~\cite{gaus13} using the DFTB+ software package.~\cite{hourahine20}

Repulsive energies and forces as given by eq.~\eqref{eq:Erep_Frep} have been obtained for a large and diverse set of organic molecules. The basis for this dataset is the QM7-X database,~\cite{qm7x} which contains molecular structures and quantum mechanical properties of small organic molecules at the level of PBE0-DFT. For the purpose of this work, we have selected molecules containing the elements \{C, N, O, H\}, but the presented methodology is easily extended to other molecular compositions. In addition to equilibrium structures, QM7-X also features 100 non-equilibrium conformations per molecule. In this work, the QM7-X database has been further extended to cover less well-represented regions of configurational space, such as select pairwise O--O and H--H distances. For further information on the augmenting structures and a short analysis of the coverage of chemical space, see Supporting Information. All presented models will be based on this final set containing $\sim$4.1~million molecules. The distribution of the final target repulsive energies and forces is shown in Fig.~\ref{fig1}.

While studying a series of ML approaches for predicting repulsive energies (see Supporting Information), the focus of this work is on employing a DTNN architecture for predicting repulsive potentials for DFTB. Hence, forming the DFTB-NN\textsubscript{rep} framework as described in Fig.~\ref{fig1}. In particular, we used the recently developed deep learning toolbox SchNetPack.\cite{Schuett2017,Schuett2018, Schuett2019} The SchNet architecture is based on atom-wise representations of molecular properties directly inferred from atomistic structures. In the layers of the DTNN, atoms are represented by a tuple of features $x_i^{(l)} \in \mathds{R}^{D}$, where $D$ is the dimension of the feature space and $l$ denotes the layer. Interactions between atoms are modeled by a series of in total $T$-times refined pairwise interactions between all $x_i^{(l)}$ within a certain cutoff, which gradually introduces information about the chemical environment (\textit{i.e.}, complex many-body terms). This procedure is carried out by using continuous-filter convolution layers with filter-generating networks. The final prediction is obtained after atom-wise updates of the feature representation and pooling of the resulting atom-wise property. Besides generating reliable DTNN-models for energy predictions, SchNet has been also proven to provide energy-conserving force models by differentiating the energy model with respect to the atomic positions.\cite{Schuett2019} Thence, we create a global DTNN-model for predicting repulsive energies and forces based on the molecules contained in QM7-X. We used $T=3$ interaction refinements, a cutoff of 5~\r{A} and a 128-dimensional feature space. For all reported results, the NN was continuously trained with a descending learning rate from $10^{-4}$ to $10^{-6}$ and a decay factor of 0.5. The training has been performed using four Tesla P100 GPUs.
Repulsive contributions as obtained in the trained SchNet model have ultimately been combined with electronic DFTB contributions as obtained with DFTB+ via a locally modified QM/MM calculator within the Atomic Simulation Environment.\cite{Bahn2002}
For the remainder of this work, we use DFTB-PW\textsubscript{rep} to refer to conventional DFTB3 with pairwise repulsive potentials. DFTB-NN\textsubscript{rep} refers to DFTB3 with the same electronic parameterization while using repulsive energies/forces from the SchNet model. For the SchNet model trained for atomization energies, we will use the term DFTB-$\text{NN}_{\text{rep}}^{\text{(E)}}$ and when trained on forces, we will use DFTB-$\text{NN}_{\text{rep}}^{\text{(F)}}$.

\begin{figure}[bt]
 \centering
 \includegraphics[width=\linewidth]{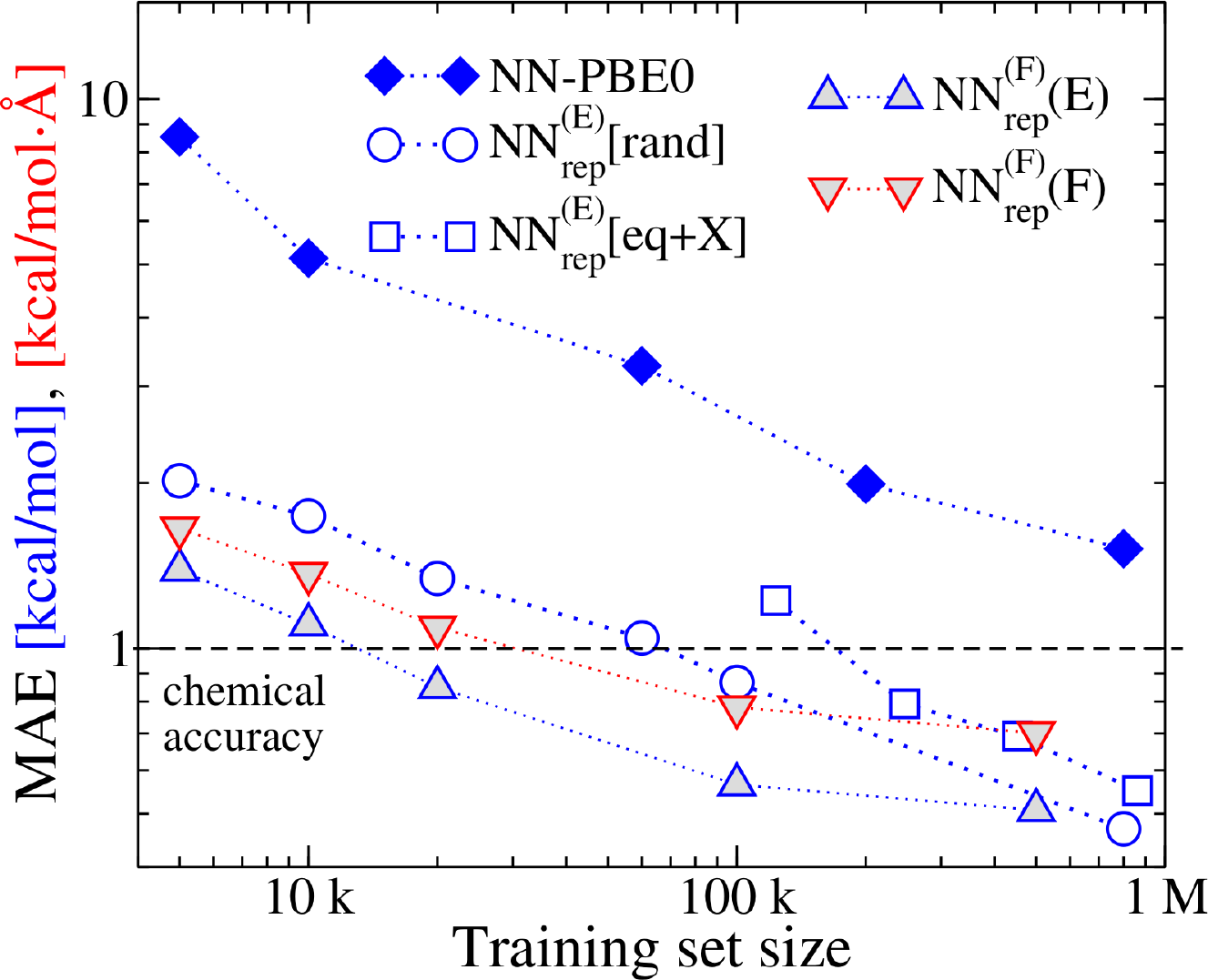}
 \caption{\textbf{Mean absolute error (MAE) as a function of training set size for the different neural network (NN) models generated in the present work.} MAEs in energies are shown for DTNN-models trained on PBE0 atomization energies (``NN-PBE0'', blue full diamonds), repulsive energies with random selection of training set (``$\text{NN}_{\text{rep}}^{\text{(E)}}$[rand]'', blue empty circles), repulsive energies with equilibrium conformations plus X non-equilibrium conformations as training set (``$\text{NN}_{\text{rep}}^{\text{(E)}}$[eq+X]'', blue empty squares), and repulsive forces (``$\text{NN}_{\text{rep}}^{\text{(F)}}$(E)'', blue shaded upward triangles). Additionally, the MAE in predicting repulsive forces is shown for the $\text{NN}_{\text{rep}}^{\text{(F)}}$ model (``$\text{NN}_{\text{rep}}^{\text{(F)}}$(F)'', red shaded downward triangles). Errors computed based on the full QM7-X dataset.}
 \label{fig2}
\end{figure}

As a first step, we studied the influence of the selection of training points on the learning process of repulsive energies and forces. This is an important issue for optimizing training set size, accuracy, and the learning progress. To this end, we considered two cases: random selection of training points from the whole dataset, which is the standard procedure in ML studies, and a training set consisting of equilibrium structures together with a given number (X) of non-equilibrium structures. In the latter case, which we will refer to as eq+X in the following, non-equilibrium structures were selected at random from domains showing low, middle or high values of \textit{E}\textsubscript{rep}. The corresponding learning curves are shown in Fig.~\ref{fig2}. On a log-log scale one expects a (near-)linear trend in the learning progress, which we here can conclusively attribute only to the randomly selected training set. For eq+X, we find linear behavior only beyond $\sim$200k training instances. When considering only equilibrium structures in the training set (a total size of $\sim$41k), for instance, the MAE is 19.5~kcal/mol, which strongly deviates from a linear trend. With an increasing portion of non-equilibrium structures, eq+X then approaches linear behavior and the results obtained with randomly selected training points. The rather unexpected finding that a random selection of the training set shows a better performance than the more refined approach of eq+X demonstrates the importance of non-equilibrium conformations and that their properties can seldom be inferred from the information on equilibrium structures. The MAE at the largest training set size for the two selection algorithms are 0.47~kcal/mol (800k) for random selection and 0.55~kcal/mol (861k) for eq+X, respectively.
Regarding the required training set sizes, we would like to note that a purely data-driven, global model can be expected to require about 5$\,\cdot\,$10\textsuperscript{6} training points to cover the compositional, configurational and conformational degrees of freedom in QM7-X (see SI, Section~4). The considerable reduction observed for NN\textsubscript{rep} can be attributed to the repulsive potential covering more local interactions, the partly included physics in the SchNet model and the shared characteristics and information among individual molecular degrees of freedom.

As mentioned above, SchNetPack also allows to train DTNN-models on atomic forces. We employed the same protocol as above to develop a DTNN for the contribution of the repulsive potential to atomic forces, $\text{NN}_{\text{rep}}^{\text{(F)}}$, which also allows to predict energies since the force model is obtained by differentiating the energy model with respect to atomic positions.\cite{Schuett2019} At 500k randomly sampled training instances, the $\text{NN}_{\text{rep}}^{\text{(F)}}$ model produces a MAE of 0.51~kcal/mol and 0.70~kcal/mol$\cdot$\AA{} for energies and forces, respectively (see Fig.~\ref{fig2}). For the eq+X sampling method the corresponding MAE values are 0.55~kcal/mol and 0.74~kcal/mol$\cdot$\AA{}. While being trained on forces, $\text{NN}_{\text{rep}}^{\text{(F)}}$ thus provides energies at an accuracy comparable to $\text{NN}_{\text{rep}}^{\text{(E)}}$, yet requires a smaller training set. This can be attributed to the increased information content in the gradient domain and the improved performance of resulting models thanks to the inherent incorporation of energy conservation.~\cite{chmiela17} We found that for 89\,\% of the molecules in QM7-X the MAE in forces is lower than the threshold of 1~kcal/mol$\cdot$\AA{} (see error distribution in Fig.~S8 of the SI). This indicates that our global $\text{NN}_{\text{rep}}^{\text{(F)}}$ model yields reliable energies and forces for the manifold molecules and conformations considered in QM7-X. The present model hence extends beyond previous works as reported by Zhu~\textit{et~al.}~\cite{zhu19}, for instance, in which DFTB forces of a single molecule (glycine) have been corrected using a NN-model. All results for the combined DFTB-NN\textsubscript{rep} approach reported in the remainder of this work were obtained using the DFTB-$\text{NN}_{\text{rep}}^{\text{(F)}}$ model, or DFTB-$\text{NN}_{\text{rep}}^{\text{(E)}}$ model where explicitly noted, with randomly selected training instances. We would like to remark that better performances can be reached by further increasing the training set size as it was shown in previous ML studies. Given the overall very low MAEs with respect to PBE0-DFT, this is beyond the scope of the current work, however.

\begin{figure}[t!]
 \centering
 \includegraphics[width=\linewidth]{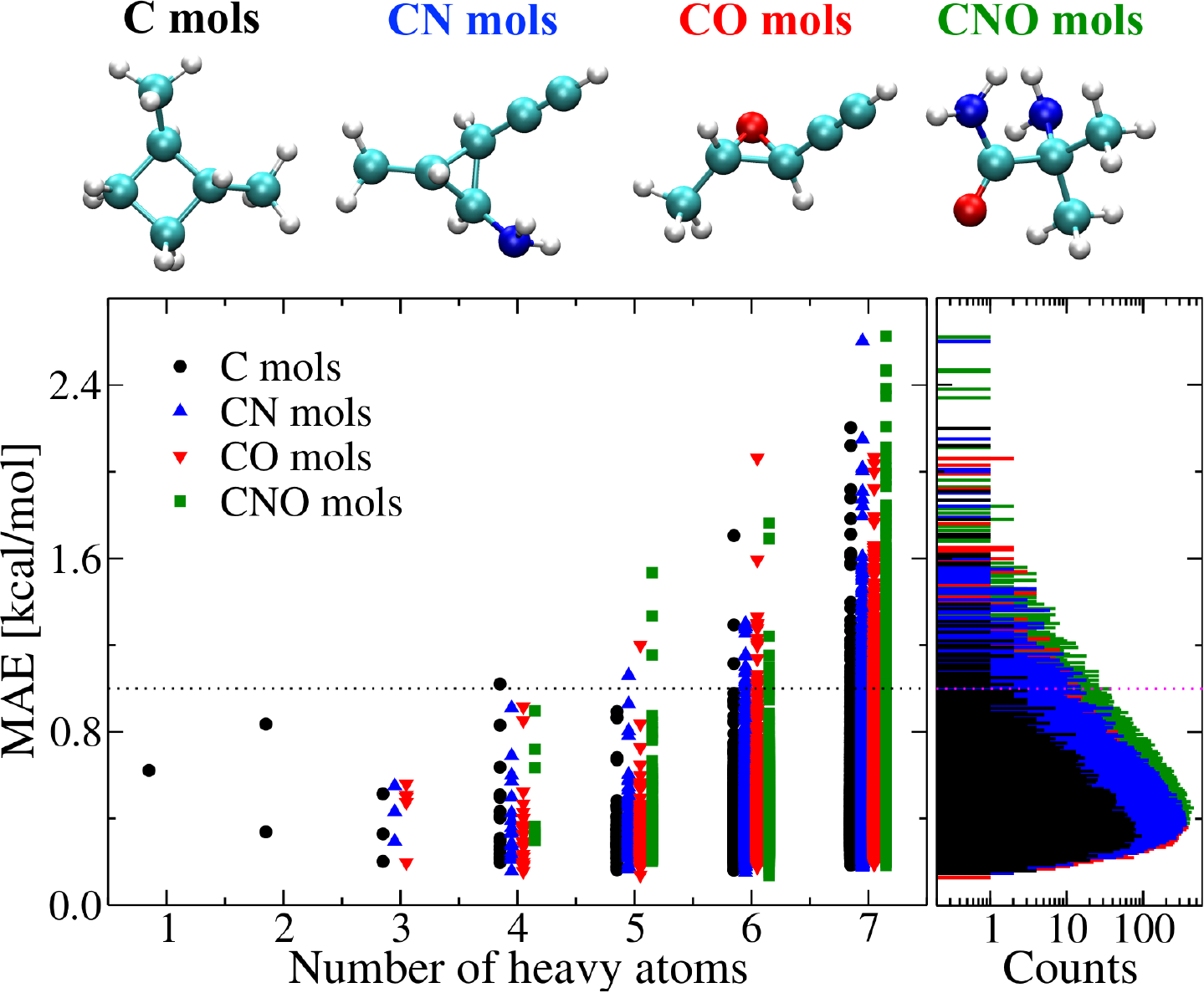}
 \caption{\textbf{Analysis of the mean absolute error (MAE) in atomization energies.} \textit{Left:} The MAE in atomization energies per molecule is shown as a function of the number of non-hydrogen atoms contained. The MAE is further separated into molecules containing only \{C,H\} (black circles), \{C,H,N\} (blue triangles up), \{C,O,H\} (red triangles down), and \{C,N,O,H\} (green squares). \textit{Right:} Distribution of MAE color-coded according to contained non-hydrogen atoms.}
 \label{fig3}
 \end{figure} 

To further analyze the performance of our NN\textsubscript{rep} models, we have split the QM7-X set into four different subgroups depending on their composition: molecules containing only the element combinations \{C,H\}, \{C,N,H\}, \{C,O,H\}, or \{C,N,O,H\} and as a function of the number of non-hydrogen atoms (see Fig.~\ref{fig3}). We then computed the MAE in atomization energies for each equilibrium molecule and its respective non-equilibrium conformations using DFTB-$\text{NN}_{\text{rep}}^{\text{(F)}}$. As a rule of thumb, the atomization energies for molecules with fewer non-hydrogen atoms and correspondingly smaller size are predicted with higher accuracy. This can mainly be attributed to the higher number of degrees of freedom in larger molecules, which increases the complexity for describing their conformational space especially in regard to strongly distorted molecules. Overall, however, 88\,\% of the molecules are predicted within an error of 1~kcal/mol, which confirms the very good performance of DFTB-$\text{NN}_{\text{rep}}^{\text{(F)}}$.
In particular, almost all hydrocarbons show very low errors thanks to the extensive sampling of C--C and C--H as exemplified for the corresponding two-, three-, and four-body combinations occurring in the training set (see Fig.~S3-5 of the SI). Consequently, the least represented molecules, which at the same time show the highest configurational and conformational complexity (\textit{e.g.}, formed by \{C,N,O,H\}) also show the largest errors due to the limited sampling. A more balanced sampling of the QM7-X reference can therefore further limit the obtained MAEs and is subject to ongoing investigations.

As the atomization energy of molecules, $E_{\rm at}$, is a widely discussed topic for general ML potentials, we have compared the results of DFTB-NN\textsubscript{rep} to a DTNN-model trained on the full PBE0 atomization energy, which we will refer to as NN-PBE0. To assure a meaningful and fair comparison, we have employed the same training sets as for NN\textsubscript{rep}. The learning curve of the NN-PBE0 model is plotted in Fig.~\ref{fig2}. NN-PBE0 follows a (near-)linear learning progress in log-scale, but shows a substantially larger MAE than NN\textsubscript{rep} and cannot overcome the threshold of 1~kcal/mol within the considered training set sizes. The minimum MAE at 800k training instances amounts to 1.52~kcal/mol. The learning progress of NN-PBE0, as characterized by the slope of the corresponding learning curve, is comparable to the one of NN\textsubscript{rep}, but thanks to the large quantum-mechanical prior in form of the electronic DFTB energy, NN\textsubscript{rep} is able to reach an MAE below 1~kcal/mol at much smaller training set sizes.
The MAE of 0.51~kcal/mol found for DFTB-NN\textsubscript{rep} also is a considerable improvement when compared to the results reported for the prediction of atomization energies using DFTB together with ML-based generalized pair-potentials ($\sim$2.6~kcal/mol, QM9 dataset).\cite{kranz18}

\begin{figure}[t!]
 \centering
 \includegraphics[width=\linewidth]{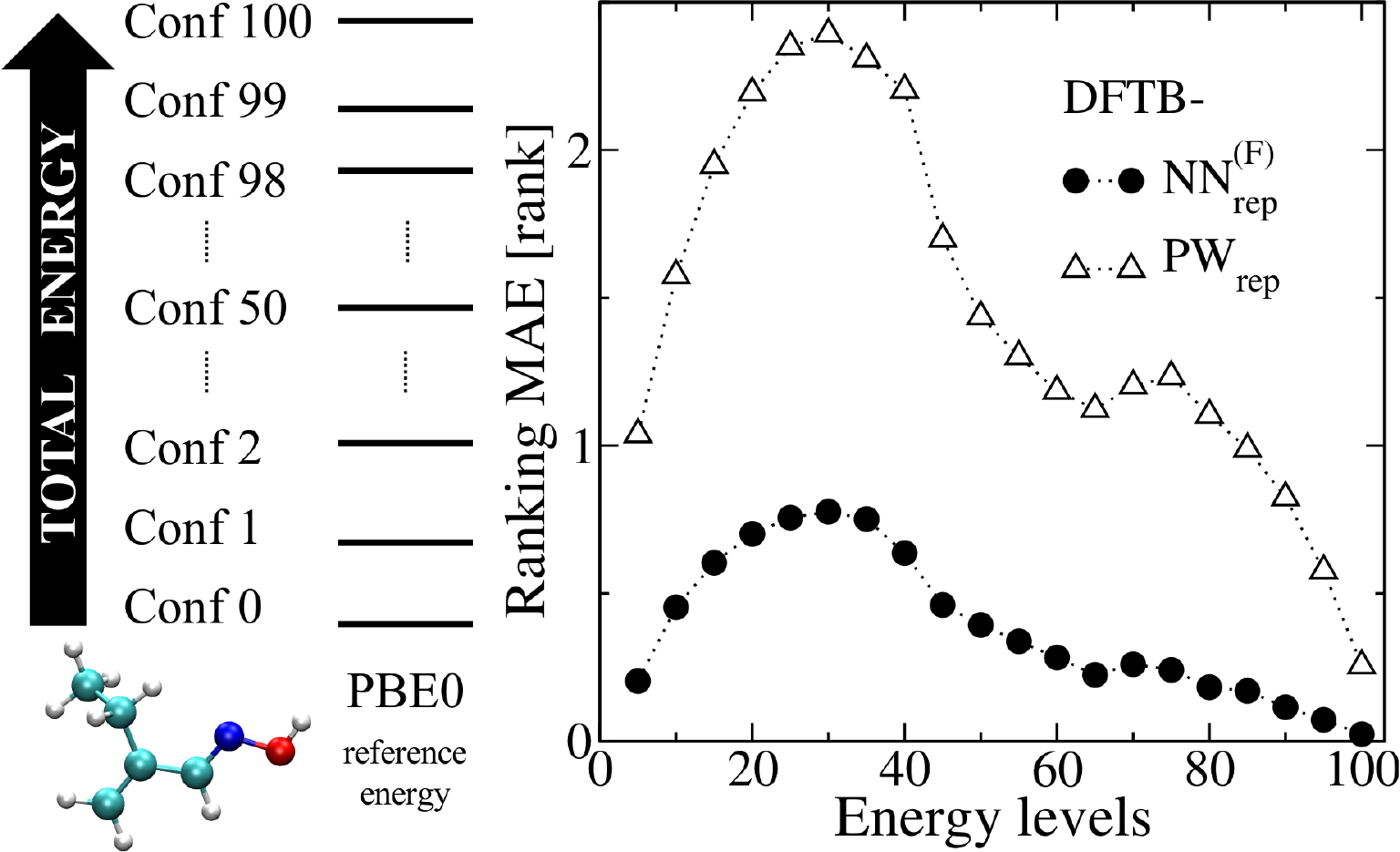}
 \caption{\textbf{Performance in predicting relative energy rankings:} Mean absolute error (MAE) in the predicted rank in comparison to PBE0 for all 100 conformations of each molecule in the QM7-X database in windows of 5 non-equilibrium conformations. Comparing relative stability rankings as obtained with DFTB-$\text{NN}_{\text{rep}}^{\text{(F)}}$ ($\text{NN}_{\text{rep}}^{\text{(F)}}$, filled circles) and conventional DFTB (PW\textsubscript{rep}, empty triangles).}
 \label{fig4}
 \end{figure}

Besides the (mean) absolute error for each conformation, the relative stabilities of each conformer represent an important and interesting performance measure. For instance, predicting the correct energetic ordering of different structures is of utmost importance in computational crystal structure prediction. To analyze the performance in predicting energy rankings, we have first determined the energetic ordering of all conformers for each molecule as given by PBE0. Based on this ordering, we can assign a rank to every structure. We then fix the ordering of the conformers and evaluate the corresponding ranks using DFTB-$\text{NN}_{\text{rep}}^{\text{(F)}}$ and conventional DFTB-PW\textsubscript{rep}. The difference between the sequences of ranks finally serves as our error measure in predicting relative stability rankings. Fig.~\ref{fig4} shows the MAE in these rank sequences with respect to the original PBE0 sequence averaged over all molecules in QM7-X. The ranking MAE is thereby further subdivided into small windows composed of five energy levels. It is evident that DFTB-$\text{NN}_{\text{rep}}^{\text{(F)}}$ (full circles) provides a much more accurate energetic ordering of the various conformers than DFTB-PW\textsubscript{rep} (empty triangles). In this regard, it has been shown previously that including relative energies in the fitting process of conventional DFTB-PW\textsubscript{rep} can provide improvements.~\cite{aguirre20} 
The general behavior observed in the ranking MAEs can thereby be explained by the energy level density per kcal/mol for each window. The more densely the conformers populate the energy spectrum, the easier can already small errors in the energy prediction cause a reordering of energetic ranks, whereas the ordering of well-separated energy levels is typically well preserved.

We have further studied isomerizations of diverse neutral molecules containing the elements C, H, N, and O. The so-called ISO34 dataset~\cite{repasky02,sattelmeyer06,grimme07} and the considered isomers are detailed in Ref.\citenum{grimme07}. The experimental (and selectively computationally refined~\cite{grimme07}) reference values of the corresponding isomerization energies have been widely used for benchmarking semi-empirical methods including DFTB.\cite{gruden17} We have tested the performance of our DFTB-NN\textsubscript{rep} models as well as for PBE0-DFT, NN-PBE0 and conventional DFTB-PW\textsubscript{rep} in predicting the isomerization energies as summarized in Table~\ref{Tab:Eiso_ISO34}. A detailed list for each reaction can be found in Table~S1 of the SI. One can see that the accuracy of DFTB-NN\textsubscript{rep} is much better than for NN-PBE0 and DFTB-PW\textsubscript{rep} with or without van der Waals correction and is in fact close to the one obtained at the level of PBE0. Computationally, however, the DFTB-NN\textsubscript{rep} model is much less expensive. For instance, PBE0 required 2 CPU hours (using 28 threads on 28 physical cores) for computing the isomerization energies of the full ISO34 dataset, while the DFTB-NN\textsubscript{rep} calculations only take 2 seconds (on one GPU).
\begin{table}[!t]
  \caption{\textbf{Summary of performance in predicting isomerization energies in comparison to experimental and CCSD(T) reference data} (in kcal/mol) as considered in ISO34 dataset:\cite{repasky02,sattelmeyer06,grimme07} mean signed error (MSE), mean absolute error (MAE) and root-mean-square error (RMSE).}
  \label{Tab:Eiso_ISO34} 
  {\centering
  \begin{tabular}{p{3.25cm} S[table-format=2.3] S[table-format=2.3] S[table-format=2.3]}
    \toprule
    Method & \multicolumn{1}{C{1.4cm}}{MSE} & \multicolumn{1}{C{1.4cm}}{MAE} & \multicolumn{1}{C{1.4cm}}{RMSE} \\
    \midrule
    PBE0-DFT       & -0.19 & 1.82 & 2.48 \\[0.5ex]
    NN-PBE0        &  2.21 & 5.85 & 11.51 \\[0.5ex]
    DFTB-$\text{NN}_{\text{rep}}^{\text{(F)}}$ & -0.71 & 2.21 & 3.30 \\[0.5ex]
    DFTB-PW\textsubscript{rep}  &  1.33 & 3.57 & 5.05 \\[0.5ex]
    DFTB-PW\textsubscript{rep}+D3$^a$& 1.4 & 3.4 & 4.9 \\
    \bottomrule\\[-2ex]
    \multicolumn{3}{l}{$^a\,$DFTB+D3 data taken from Ref.\citenum{gruden17}}
  \end{tabular}}
\end{table} 
In addition to the isomerization energies, we have investigated the performance of our DFTB-$\text{NN}_{\text{rep}}^{\text{(F)}}$ model in predicting the equilibrium structures and vibrational mode frequencies of the isomers in ISO34. Table~\ref{Tab:geom-vib_ISO34} compares the root mean square deviation of the optimized structures (RMSD\textsubscript{s}) and mean absolute error in vibrational frequencies (MAE\textsubscript{$\omega$}) with respect to PBE0 for DFTB-PW\textsubscript{rep} and our DFTB-$\text{NN}_{\text{rep}}^{\text{(F)}}$ model. For optimized structures, both approaches perform similar and well on average. However, DFTB-$\text{NN}_{\text{rep}}^{\text{(F)}}$ displays a better performance for a wider range of substrates and products, see Table~S2. The only cases in which DFTB-PW\textsubscript{rep} provides better results than DFTB-$\text{NN}_{\text{rep}}^{\text{(F)}}$ correspond to more complex structures composed of eight non-hydrogen atoms or unseen functional groups. In the case of vibrational frequencies, DFTB-$\text{NN}_{\text{rep}}^{\text{(F)}}$ provides a much superior description for all isomers without exception (see Table~S3). It is worth mentioning that vibrational calculations using DFTB-$\text{NN}_{\text{rep}}^{\text{(F)}}$ took $\sim$1.5 hours (on one GPU), while the PBE0-DFT calculations required $\sim$295 CPU hours (using 28 threads on 28 physical cores). Previous studies have already shown that conventional DFTB parameterizations seldom allow for an accurate prediction of energetic, structural, and vibrational properties at the same time. As a result, special-purpose parameterizations have been devised for vibrational analysis, for instance.~\cite{gaus13,hourahine20} The presented DFTB-$\text{NN}_{\text{rep}}^{\text{(F)}}$ framework, on the other side, indeed does enable accurate predictions of energetic (\textit{cf.} Table~\ref{Tab:Eiso_ISO34}), as well as structural and vibrational properties. This is essential for, \textit{e.g.}, consistent and seamless calculations of vibrational spectra, free energies or other thermodynamic and transport properties.
\begin{table}[hbt!]
  \caption{\textbf{Structural root mean square deviation of equilibrium geometries} (RMSD\textsubscript{s}) and \textbf{mean absolute error of vibrational frequencies} (MAE\textsubscript{$\omega$}) as predicted by DFTB-PW\textsubscript{rep} and DFTB-$\text{NN}_{\text{rep}}^{\text{(F)}}$. All errors are reported for the complete set of molecules in ISO34, sub-categorized according to the contained non-hydrogen atoms, and obtained in reference to PBE0 results. See SI for further information.}
  \label{Tab:geom-vib_ISO34} 
  \centering
  \begin{tabular}{p{1.8cm} C{1.3cm} C{1.3cm} p{0.1cm} C{1.3cm} C{1.3cm}}
    \toprule
     & \multicolumn{2}{c}{RMSD\textsubscript{s} [\AA{}]} & & \multicolumn{2}{c}{MAE\textsubscript{$\omega$} [cm$^{-1}$]} \\[0.5ex]
    DFTB- & PW\textsubscript{rep} & $\text{NN}_{\text{rep}}^{\text{(F)}}$ & & PW\textsubscript{rep} & $\text{NN}_{\text{rep}}^{\text{(F)}}$ \\
    \midrule
    C mols   & 0.03 & 0.02 & & 49.9 & 7.41 \\
    CN mols  & 0.05 & 0.02 & & 49.8 & 8.38 \\
    CO mols  & 0.06 & 0.02 & & 50.2 & 6.72 \\
    CNO mols & 0.24 & 0.23 & & 51.9 & 6.68 \\
    \bottomrule
  \end{tabular}
\end{table}

\begin{figure*}[!tb]
    \centering
    \includegraphics[width=0.95\linewidth]{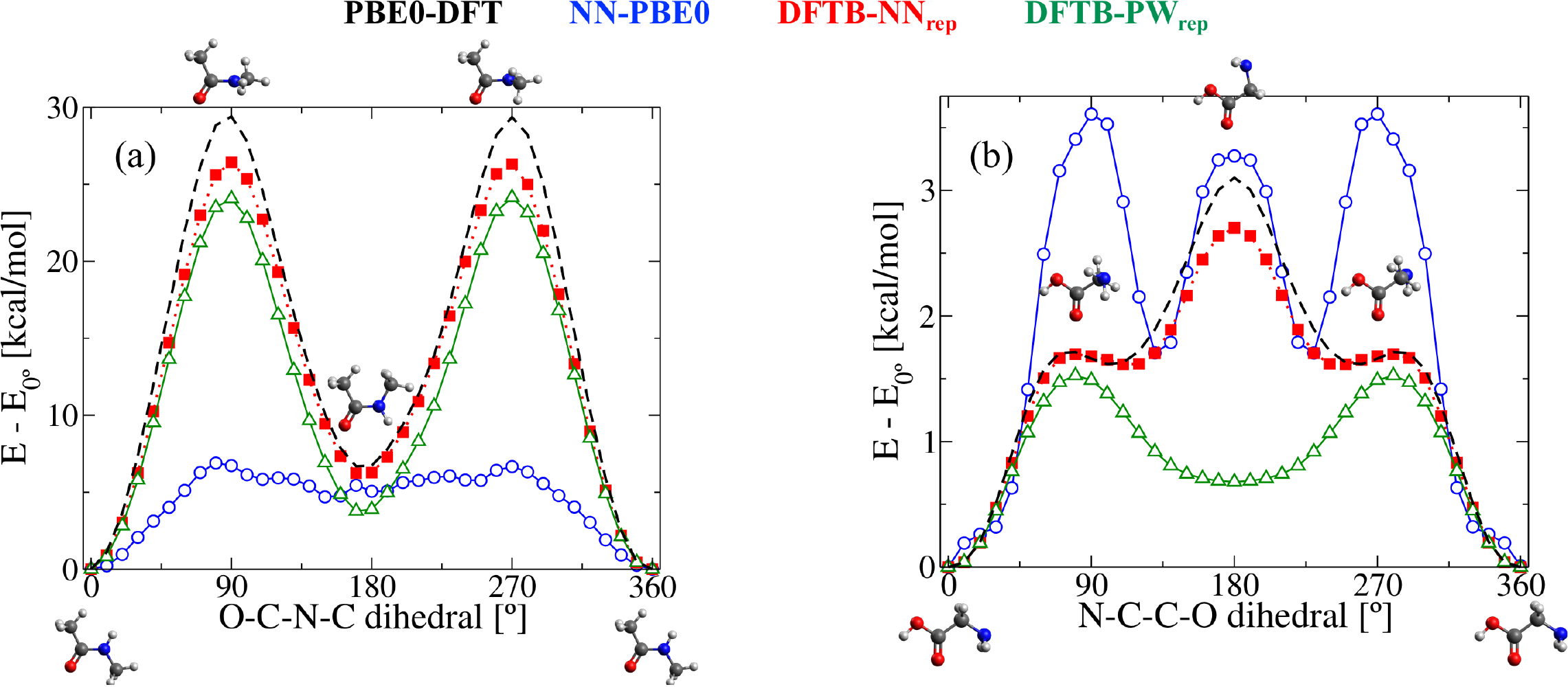}
    \caption{\textbf{Potential energy profiles upon dihedral rotation.} PBE0-DFT, NN-PBE0, DFTB-NN\textsubscript{rep} and DFTB-PW\textsubscript{rep} profiles for rotating a) the O--C--N--C dihedral of a peptide bond and b) the N--C--C--O dihedral in glycine.}
    \label{fig5}
\end{figure*}
To finally highlight the advantages of our DFTB-NN\textsubscript{rep} formalism and its good transferability among small organic molecules, we have also studied the prediction of dihedral rotation profiles. As test cases, we chose a simple peptide bond ({O--C--N--C} dihedral angle) and glycine {(N--C--C--O)}. The corresponding potential energy profiles are shown in Fig.~\ref{fig5}. In the case of the peptide bond, DFTB-PW\textsubscript{rep} (green triangles) underestimates rotational barriers by 5.4~kcal/mol and overestimates the relative stability of the meta-stable intermediate by 2.8~kcal/mol in comparison to PBE0 (dashed black). DFTB-NN\textsubscript{rep} (red squares) considerably improves the performance only underestimating the barriers by 2.9~kcal/mol. For the {N--C--C--O} dihedral of glycine, finally, traditional DFTB-PW\textsubscript{rep} even misses qualitative features of the rotational profile: Higher-level reference calculations (PBE-, PBE0- and B3LYP-DFT as well as MP2) predict two low-energy barriers at 90 and 270$^\circ$ and a main barrier of $\sim$3~kcal/mol at a dihedral of 180$^\circ$. DFTB in conjunction with the traditional pairwise repulsion completely misses this most relevant barrier and instead predicts a shallow minimum between 90 and 270$^\circ$. Our DFTB-NN\textsubscript{rep} model is able to correct for this shortcoming and provides very good agreement with the reference results. This is particularly encouraging considering that neither the glycine molecule nor any rotational profiles were part of the training set. 
For comparison, we also computed dihedral energy profiles as predicted by NN-PBE0 (blue circles). As evident from Fig.~\ref{fig5}, the DTNN directly trained on PBE0-DFT data is unable to predict meaningful rotational profiles. Together with the poor performance for the ISO34 set (\textit{vide supra}), we can conclude that NN-PBE0 has a very limited validity outside the immediate scope of its training set. The much increased transferability of our DFTB-NN\textsubscript{rep} approach can be attributed to the more local character of \textit{E}\textsubscript{rep} and further motivates the exploration of combining (approximate) quantum-mechanical Hamiltonians with ML.

In conclusion, this work presents a successful and promising example of combining semi-empirical electronic structure methods with ML-based approaches to localized many-body interactions. We introduce a NN-based approach for the prediction of improved, beyond-pairwise repulsive potentials for DFTB. The overall accuracy and reliability of DFTB-NN\textsubscript{rep} was demonstrated for atomization energies and relative stability rankings across the vast configurational and conformational diversity of small organic molecules considered in the QM7-X dataset. The applicability outside the scope of the training set has been highlighted for isomerization energies, structural and vibrational characteristics of molecules as contained in the ISO34 dataset and dihedral rotation profiles of small organic molecules. As such, the present work extends beyond previous works,\cite{kranz18,zhu19} which typically involve only one test set, cover either configurational \textit{or} conformational diversity, or study only a very limited set of (energetic) properties. Since the main target of the presented methodology is the repulsive component of DFTB, the performance in terms of electronic properties remains largely untouched and can be optimized separately. It has recently been proposed to employ NN-based approaches for such optimizations as well.~\cite{li18} Finally, we would like to emphasize that the presented protocol can be equally well paired with other electronic parameterizations and for arbitrary target systems. Investigations of the interplay of electronic parametrization and the learning process are subject to ongoing work.

Further improvements of DFTB-NN\textsubscript{rep} concern the intricate task of efficient and sufficient sampling, which is known to limit the validity of ML models for increasing complexity and structural flexibility. This is one of the key aspects for the increasing MAEs when going to larger molecules or higher-energy non-equilibrium conformations (see Fig.~\ref{fig3} and S6). Another shortcoming of DFTB-NN\textsubscript{rep} is its limited scalability toward system sizes well beyond the training set, especially when involving completely different molecular graphs. While showing improved transferability compared to NN-models for the total PBE0 energy, NN\textsubscript{rep} still does not extrapolate well towards larger scales or completely new structure motifs. In this sense, we are confident that an improved selection of training instances will alleviate such issues. Further analysis of local environments, as shortly presented in Fig.~S4 and S5 of the SI, can thereby provide the necessary information for a more efficient and balanced sampling.
Besides potential shortcomings in the sampling of training instances, the above issue can partly be attributed to NN\textsubscript{rep} not fully exploiting the more localized character of the target quantity. One promising remedy in that regard is to limit the descriptor space and enforce a more local model, while retaining a beyond-pairwise description. This can provide an optimal trade off between accuracy and robustness with respect to spurious effects of global structure motifs on the descriptor. A more local model will also require considerably less training instances than the global DTNN employed here. In this context, we would like to emphasize that the total number of degrees of freedom in QM7-X is about $2\cdot 10^{6}$ and a global, data-driven model is expected to require a training set of $\sim$5$\,\cdot\,$10\textsuperscript{6}. Yet, the space spanned by the local chemical environments in the dataset can be estimated to involve only $\sim$120 degrees of freedom requiring about 5000 unique samples (see SI, Section~4). The first and foremost step forward is thus the ongoing development of ML models specifically designed for localized many-body repulsive potentials in DFTB. Future studies following up on the ideas outlined above will help to further advance the applicability of DFTB-NN\textsubscript{rep} to treat systems and phenomena of larger physical and chemical complexity.

The authors gratefully thank Farnaz Heidar-Zadeh for helpful discussions during the early design of this work and Kristof T. Sch\"{u}tt and Michael Gastegger for advice on performing and optimizing applications within the SchNetPack toolbox.
MS acknowledges financial support from the Fonds National de la Recherche Luxembourg (AFR PhD grant CNDTEC).
LMS and AT were supported by the European Research Council (ERC-CoG BeStMo).

\bibliography{arXiv-references}

\end{document}